\documentclass[aps,prl,twocolumn,showpacs,amsmath,amssymb,superscriptaddress]{revtex4}
\usepackage{graphicx}
\begin{document}

\title{Quantum transport thermometry for electrons in graphene}
\author{K. Kechedzhi}
\affiliation{Department of Physics, Lancaster University,
Lancaster, LA1 4YB, UK}
\author{D. W. Horsell}
\author{F. V. Tikhonenko}
\author{A.~K.~Savchenko}
\author{R. V. Gorbachev}
\affiliation{School of Physics, University of Exeter, Stocker Road,
Exeter, EX4 4QL, UK}
 \author{I. V. Lerner}
\affiliation{School of Physics and Astronomy, University of
Birmingham, Egbaston, Birmingham B15 2TT, UK}
\author{V. I. Fal'ko }
\affiliation{Department of Physics, Lancaster University,
Lancaster, LA1 4YB, UK}

\begin{abstract}
We  propose a method of measuring the electron temperature $T_e$ in
mesoscopic conductors and demonstrate experimentally its
applicability to micron-size graphene devices in the linear-response regime ($T_e\approx T$, the bath temperature). The method  can
be {especially useful} in case of overheating, $T_e>T$. It is based on analysis of the
correlation function of mesoscopic conductance fluctuations.
Although the fluctuation amplitude strongly depends on
the details of electron scattering in graphene, we show that $T_e$ extracted from the
correlation function is insensitive to these details.
\end{abstract}


\maketitle
\bibliographystyle{myprb}

Graphene is an atomically thin graphite layer~\cite{Wallace,McClure} recently used in
field-effect transistors~\cite{NovGeim}. In graphene-based semiconductor devices
phonons are poorly coupled to the environment since the mass of
carbon atoms is typically smaller than that of atoms in the
underlying substrate, making the overheating of graphene structures
a likely event at high currents. This raises {a} question of
how to measure the temperature of electrons in graphene. Since
classical conductivity in graphene has a very weak temperature
dependence at low and intermediate temperatures \cite{Geim:08},
extracting the electron temperature from transport measurements
requires analyzing more subtle quantum effects. One possibility
would be to analyze the decoherence rate $\tau_{\varphi}^{-1}$ using
the weak-localization (WL) {effects in} magneto-resistance \cite{AAG,Marcus:93}.
However, it
was shown theoretically \cite{McCann:06} 
and confirmed experimentally \cite{Sav:08} that the WL in graphene
reveals itself in a rather complicated way due to the influence of
inter-valley scattering and the disorder which breaks the sublattice
symmetry. Thus WL does not offer an easy way of measuring the electron temperature $T_e$. Another
possibility would be to exploit the temperature dependence of the amplitude
of universal conductance fluctuations (UCF) \cite{AAG,AlKh:85,LSF}.
Unfortunately, a quantitative implementation of such analysis is
hindered by the necessity to both account for the temperature
dependence of $\tau_{\varphi}^{-1}$
and attribute a definite symmetry class to a particular
graphene-based device \cite{Kechedzhi:08,Sav:08,Khar,Wurm}. However, it has been noticed that
the correlation functions
of random UCF dependence on magnetic field $B$ and the Fermi energy $\varepsilon _\text{F}$ provide
useful information about {subtle} spectral characteristics of a
disordered conductor~\cite{Haug1}.
\begin{figure}\vspace*{1mm}
 \includegraphics[width=6.5cm]{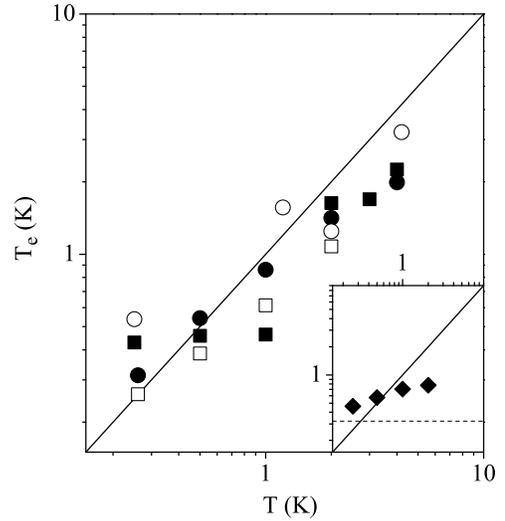}
\caption{The electronic temperature
$T_e$, determined via Eq.~(\ref{dEc1}), as a function of the bath temperature, $T$, for four
graphene flakes (F1: open circles,
F2: open squares, B1: filled circles, B2: filled squares,  see Table \ref{t1}).  The inset shows $\Delta_c/2.7$ for sample
D  with $L_T\sim L_x$, i.e.\ not in the regime (\ref{L});  the dashed horizontal line
shows the Thouless energy, $h/\tau_D$. }\label{fig:1}
\end{figure}

In this Letter we propose a method of correlation function
thermometry of mesoscopic conductors and demonstrate experimentally
its applicability to graphene-based micron-size devices. Although  the UCF in graphene differ in detail \cite{Kechedzhi:08,Khar,Wurm} from those in other mesoscopic conductors, the proposed method is robust and independent of such details. It
is based on the analysis of the normalized correlation function $F_\text{n}\left( \Delta
 \right)$ of conductance
fluctuations, $\delta G=G-\left<G\right>$, as a function of the Fermi
energy (using a wide range of the UCF magnetofingerprints for statistically representative averaging \cite{anglbrckt}):
    \begin{align}\label{F1}
F_\text{n}(\Delta)&\equiv\frac{ F(\Delta)}{ F(0)}\,,&        F(\Delta)& \equiv \bigl<\!\bigl< \delta G
    (\varepsilon_{\text F}) \delta G(\varepsilon_{\text F} +
    \Delta)\bigr>\!\bigr>\, .
    \end{align}
This function depends on the electron temperature $T_e$ which governs the thermal broadening of the Fermi
distribution. We show that for a quasi-1D wire
the width $\Delta_c$ of $F_\text{n}$ (taken at the half-maximum) is  defined by $T_e$:
    \begin{align}\label{dEc1}
    \Delta_c \approx 2.7 k_{\text B}T_e\,.
    \end{align}
 This expression allows one to determine  $T_e$ by extracting $\Delta_c$ from measuring the correlation function (\ref{F1}).  The result is valid with  accuracy of about $10\%$ {provided that the following  conditions are fulfilled:
\begin{subequations}\label{L}\begin{align}\label{LT}
    L_T&\ll \min(L_\varphi,\,L_x) \,,\\
 L_y&\ll \min(L_\varphi,\,L_x)\,,\label{Ly}
\end{align}\end{subequations}
where $L_T\equiv\sqrt{\hbar D/k_\text{B}T_e }$ is the thermal smearing length and $L_\varphi \equiv\sqrt{D\tau_\varphi }$ is the dephasing length,  $D$ is the diffusion coefficient. Inequality (\ref{LT}) defines an experimentally relevant ``high-T" regime (which may extend well below 1K) and is paramount for the method to work. Inequality (\ref{Ly}) is less demanding: one can use Eq.~(\ref{dEc1}) to determine $T_e$ also for a $2$D sample with  aspect ratio $a\equiv
L_x/L_y \sim 1$  up to $L_\varphi/L_y\sim1 $}, albeit with lesser accuracy \cite{Tdistr}.

We have experimentally tested the method in four graphene flakes in the regime (\ref{L}), using low-cur\-rent measurements to avoid electron overheating. Then $T_e $ should coincide with the bath  temperature $T$. We show the results in Fig.~\ref{fig:1}, where $T_e$ xtracted from Eq.~(\ref{dEc1}) and the directly measured $T$ are, indeed, in  good agreement.

{Below we first derive our main result, Eq.~(\ref{dEc1}), then proceed with its numerical testing and finally discuss experimental results in more detail.}

\begin{figure}[t]
\includegraphics[width=7cm]{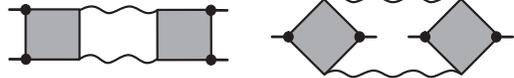}
\caption{The diagrams which contribute to the main order in
the diagrammatic expansion of the conductivity-conductivity
correlation function. The  wavy lines stand
for the propagators of the diffusion modes describing motion at length scales $\gg \ell$;
shaded blocks stand for  Hikami boxes, {describing motion at length scales $\sim\ell$}.} \label{fig:2}
\end{figure}

{The brackets $\left\langle \left\langle \dots \right\rangle  \right\rangle $ in Eq.~(\ref{F1}) stand for both the ensemble and thermal averaging. This equation can be explicitly represented} \cite{AlKh:85,LSF} as the following convolution:
 \begin{gather}
     F(\Delta)
    =  \int
    \textrm{d}\varepsilon
    K(\varepsilon,\Delta)\mathcal{F}(\varepsilon) \,,\notag\\[-8pt] \label{Kc} \\[-8pt]
    K(\varepsilon,\Delta) \!=\! \left( \frac{4e^2}{h} \right)^{\!2}\!\!\!
    \int \textrm{d}E f'(E,\varepsilon_{\text F})
    f'(E + \varepsilon,\varepsilon_{\text F}+\Delta)\,.\notag
    \end{gather}
{Here $\mathcal{F}(\varepsilon)\equiv\left\langle \delta G(E)\,\delta G (E+\varepsilon ) \right\rangle $ is the ensemble-aver\-aged correlator of conductance fluctuations at different energies and  $K$ is the thermal
broadening factor},
where the
energy derivative of the Fermi distribution function is $f'(E,\varepsilon_{\text F}) = -1/(4k_{\text B}T_e)
\cosh^{-2}[ (E-\varepsilon_{\text F})/(2k_{\text B}T_e)]$. {The standard diagrams for  $\mathcal{F}(\varepsilon )$ in the lowest order in ${\hbar/(p_{\text F} \ell) \ll 1}$ ($\ell$ is the electron mean-free path)  are  shown in
Fig.~\ref{fig:2}.}  {Structurally, they coincide with the diagrams
describing mesoscopic fluctuations in usual conductors~\cite{AlKh:85,diags} but the so-called Hikami boxes are different \cite{Kechedzhi:08,Khar,Wurm}  because of graphene-specific features: the linear dispersion law, chirality of the carriers and valley degeneracy. These features, being paramount for a quantitative description of the UCF in graphene  \cite{Kechedzhi:08,Khar,Wurm}, have no impact on calculating ${F}_\text{n}$. We will show this by analyzing first a narrow graphene wire, Eq.~(\ref{Ly}), with a strong inter-valley scattering (induced, e.g., by atomically sharp disorder). Then the part of the correlator $\mathcal{F}$} which contributes to $F_\text{n}$ can be written via
the ``valley-singlet'' diffusion propagators  $\mathcal{D}_{nm}$,
neglecting the ``valley-triplet'' modes (see Refs.~\cite{McCann:06,Kechedzhi:08} for the appropriate classification of the diffusion modes in graphene):
   \begin{subequations}\label{FD} \begin{align}
    \mathcal{F} (\varepsilon) &= \sum_{n,m}
      \left(\left| \mathcal{D}_{nm}
    \right|^2%
    + \frac{1}{2}\Re \left[ \mathcal{D}_{nm}
    \right]^2\right), \label{eq:curlF}
    \\\label{DM}
    \mathcal{D}_{nm}&\equiv \left[-\frac{\textrm{i}}{\hbar} \varepsilon
    \tau _D  + (\pi n)^2
    +  \left(a \pi m \right)^{\!2}   +
    \frac{\tau_D}{\tau_\varphi}\right]^{\!-1}\,,
    \end{align}\end{subequations}
    where $a=L_x/L_y$ and $\tau_D=L_x^2/D$.
Under the conditions (\ref{L}) the sum in Eq.~(\ref{eq:curlF})
is dominated by the $m \!=\! 0$ term.
There we have taken into account only the  diffusion modes  and neglected the Cooperons. This corresponds to the regime of suppressed WL, when the magnetic flux through the area of order $L_\varphi^2$ is much bigger than the flux quantum. In this regime  a wider range of statistical data is available due to the averaging over magnetic fields \cite{anglbrckt}.

{In order to get an asymptotic analytical} expression for the correlation
function $F_n(\Delta)$ {we assume that in addition to Eq.~(\ref{L}) the dephasing is sufficiently strong,} i.e.\  $L_x\gg L_\varphi$.
Keeping only the term
with $m=0$ and performing the summation
over $n$, we arrive at
   \begin{align}\label{int}
    \mathcal{F}(\varepsilon )=\frac{1}{2\sqrt{2} }
    \left( \frac{L_\varphi }{L_x}  \right)^{\!3}
    \frac{3t^2+t+2}{t^3\sqrt{t+1} }-\left( \frac{L_\varphi }{L_x}  \right)^{\!4}
    \frac{t^2+2}{t^4}\,,
   \end{align}
where $t \equiv  \sqrt{(\varepsilon \tau_\varphi /\hbar)^2+1}$. The
correlator $\mathcal{F}(\varepsilon )$ in Eq.~(\ref{int}) is a sharply peaked
function of $\varepsilon $ with maximum at $\varepsilon=0$ and width
$\hbar/\tau_\varphi $. In contrast, the thermal broadening factor
$K(\varepsilon,\Delta)$ in Eq.~(\ref{Kc}) has a broad
peak around $\varepsilon=\Delta$ with the width of the order of
$k_{\text B}T_e \gg \hbar/\tau_{\varphi}$, {according to Eq.~(\ref{LT}).}  Therefore the integration
over $\varepsilon$ in Eq.~(\ref{Kc}) can be performed using the
mean value theorem, i.e.\ taking $K(0,\Delta)$ out of the integral.
{As a result the normalized correlation
function becomes independent of the microscopic details contained in $\mathcal{F}(\varepsilon )$}:
    \begin{gather}
    { F_\text{n}(\Delta) }   = \frac{K(0,\Delta)}{K(0,0)} =
     \frac{3\left(\theta\coth\theta - 1\right)}
     {\sinh^{2}\theta}
     ,\;\; \theta\equiv \frac{\Delta}{2k_{\text B}T_e}\,.\label{F}
    \end{gather}
The width of this function at the half-maximum is $\theta_c=1.36$ which results in Eq.~(\ref{dEc1}). {We stress again that this result is truly universal: a precise form of $\mathcal{F}(\varepsilon )$ is irrelevant for $F_n(\Delta)$ in Eq.~(\ref{F}). The only requirement for its validity is that the function $\mathcal{F}(\varepsilon )$ in Eq.~(\ref{Kc}) is sharply peaked compared to $K(\varepsilon, \Delta)$.} {This remains valid  under the condition (\ref{LT}) for any dephasing, $L_\varphi\lesssim L_x$, and allowing for all the diffusion modes in graphene or, indeed, in any other mesoscopic disordered conductor.}

\begin{figure}\vspace*{3mm}
\includegraphics[width=7cm]{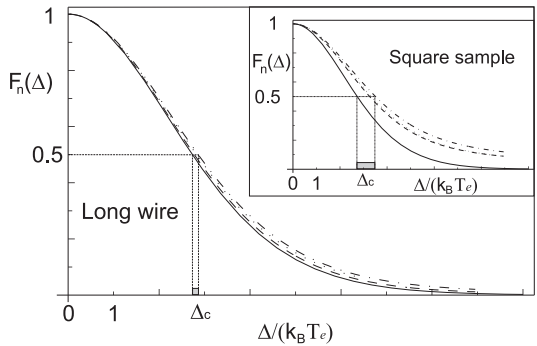}
\caption{The normalized correlation function, $F_\text{n}(\Delta)$,
  {for a wire under conditions (\ref{L}).
The solid line corresponds to the asymptotic Eq.~(\ref{F}). The
dotted, dashed, and dash-dotted lines correspond to the numerical
integration of Eqs.~(\ref{Kc}) with $\alpha\!=\!0.01, \beta\!=\!0.3;
\;\alpha\!=\!\beta\!=\!0.1; \; \alpha\!=\!0.1,\,\beta\!=\!0.3$, respectively
($\alpha\equiv L_T/L_x,\,\beta\equiv L_T/L_\varphi $).} The width at half maximum is $\Delta_c
\approx \left(2.8 \pm 0.1\right) k_{\text B}T_e$. The inset shows
$F_\text{n}(\Delta)$ for a square sample  where each line has the same
values of $\alpha$ and $\beta$.}\label{fig:3}
\end{figure}

{We have checked this numerically, calculating  $F_\text{n}(\Delta)$
  for a wide range of
$T$ and $\tau_\varphi$,  $0.01\leq
L_T/L_\varphi\leq 0.3$ with $ L_\varphi
\leq L_x$, and at various values of the symmetry breaking parameters. We have also considered a case of smooth disorder, taking into account the valley-triplet diffusion channels [neglected in the analytical calculations in Eqs.~(\ref{FD})--(\ref{int})]. A few representative examples are
plotted in Fig.~\ref{fig:3}. The values of
$\Delta_c$ lie within a narrow interval, $2.7  \leq
\Delta_c /k_{\text B}T_e\leq 2.9  $, close to the asymptotic value $2.7$ of Eq.~(\ref{dEc1}). This shows} that the proposed method works in the quasi-1D case with accuracy of about $10\%$.
We have also performed a similar analysis of 2D samples, calculating
$F_\text{n}(\Delta)$ for the same range of parameters as for wires.
The results plotted in the inset to Fig.~\ref{fig:3} indicate that
$\Delta_c\approx3k_\text{B} T_e$, i.e.\ {the method still can be applied albeit with  accuracy of about $25\%$.}

We have tested the feasibility of the proposed method in graphene-based structures. To this end, we have experimentally determined the width of the  correlation function (\ref{F1}) and thus  the electron temperature $T_e$, Eq.~(\ref{dEc1}). We have compared $T_e$ to the bath temperature $ T$ in
low-current measurements, i.e.\ in the regime when graphene is not overheated, and found them to be in satisfactory agreement, Fig.~\ref{fig:1}.
\begin{figure}[b]
\includegraphics[width=7.5cm]{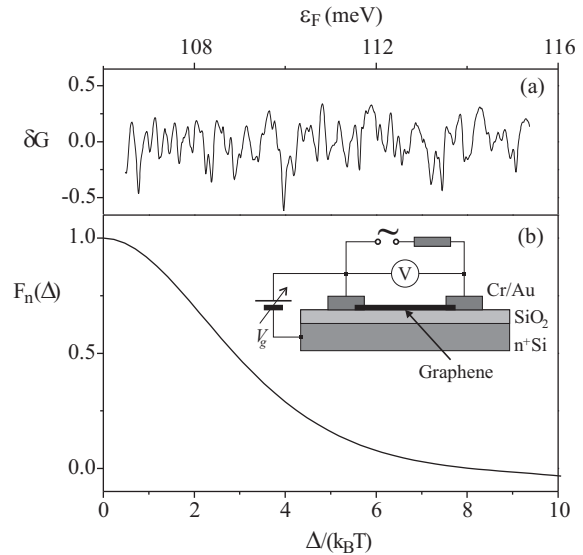}\caption{(a)
Typical fingerprint of $\delta G$ (normalized by $e^2/h$) for sample
B1 at $T=0.25$\,K and $B=90$\,mT as a function of Fermi energy (controlled by the gate voltage \cite{efermi}.) (b)
Correlation function  for sample B1 at $T=0.25\,$K. The inset
sketches the circuit used in the experiment.   The resistor has
resistance much greater than that of the graphene flake to maintain a fixed
ac current of 1\,nA.}\label{fig:4}
\end{figure}
The experimental samples used for  correlation-function
thermometry are monolayer graphene flakes created
by mechanical exfoliation \cite{Geim0} on a n$^+$Si substrate
covered by $300\,$nm of SiO$_2$. The flakes are connected
electrically by two Au/Cr contacts in the circuit shown in the inset
to Fig.~\ref{fig:4} and thermally anchored within the $^3$He pot of
a sorption-pumped cryostat. The ac current driven through the sample
was 1\,nA (at $\sim10\,$Hz). Heating by the
current was detected by measuring the mesoscopic fluctuations, and
this current was reduced until it had no effect on the
fluctuations. The Fermi energy was controlled by a gate voltage $V_g$
applied between the substrate and the flake~\cite{efermi}. Samples
characteristics are described in Table~\ref{t1}. The dephasing rate
in all samples was determined in a way similar to that in
\cite{Sav:08} from a fit of the magnetic field dependence of the
sample conductivity to the theory of weak localization in graphene
\cite{McCann:06}. For these samples  at high carrier density,
$\sim10^{12}\,$cm$^{-2}$, min$(L_x,L_\varphi)$ satisfies the conditions (\ref{L}) for the applicability of the method.

\begin{table}[t]
    \begin{tabular}{l|c|c|c|r|c}
 & $\;L_x\:$($\mu$m) & $\;L_y\:$($\mu$m) & $\;L_\varphi\:$($\mu$m) & $\ell\:$(nm) & $\; n\,$($10^{12}$cm$^{-2}$)
\\\hline
D  & 1.4 & 1.4 & 1.2 & 70\;\;\; & 1.4\\
F1 & 4.1 & 1.8 & 1.7 & 70\;\;\; & 1.4 \\
F2 & 3.8 & 1.8 & $>\!L_x$ & 120\;\;\; & 0.7 \\ B1 & 3.7 & 0.3 & $>\!L_x$ & 80\;\;\; & 0.9 \\
  B2 & 2.0 & 0.3 & $>\!L_x$ & 50\;\;\; & 0.9  \\
\end{tabular}\caption{Characteristics of the samples used for the
correlation function thermometry, Fig.~\ref{fig:1}. The values of $L_\varphi $ are given at $T=0.26$K; for the last three samples, where $L_\varphi>L_x$, it cannot be determined from the WL measurements.}\label{t1}
\end{table}

Mesoscopic fluctuations of the conductance $G$ occur in all samples
as a function of both magnetic field and Fermi energy,
Fig.~\ref{fig:4}(a), and are reproducible for the forward/backward
sweeps of $\varepsilon_F$ and $B$. The amplitude and correlation
function both depend on the bath temperature $T$ over the full temperature
range of the experiment ($0.25-20\,$K). A range of Fermi energies is
chosen such that the average resistance does not change
significantly over this range and contains sufficient number
($>100$) of fluctuations for averaging (at $T\gtrsim10\,$K such a
requirement cannot be satisfied so we restrict analysis to
$T<10\,$K). We determine $F_n(\Delta) $, Fig.~\ref{fig:4}(b),
and its width at half maximum $\Delta_c$  performing
measurements over the magnetic field range $50\,\mathrm{mT}\lesssim B\lesssim
300\,\mathrm{mT}$ \cite{anglbrckt}. It is shown in Fig.~\ref{fig:1} as a function of
the bath temperature and agrees with the theoretical relation,
Eq.~(\ref{dEc1}), with $T_e\approx T$. For comparison, the inset to
Fig.~\ref{fig:1} shows that $\Delta_c$ for sample D (which is not in
the ``high-temperature" regime of Eq.~(\ref{LT})) saturates at
$h/\tau_D$.

{In conclusion, we have proposed to use the correlation function
of mesoscopic fluctuations in disordered samples for determining the electron temperature from its width.}  {We have shown that the method
is universal and independent of microscopic details of a disordered mesoscopic sample: Eq.~(\ref{dEc1}) holds for any sample in regime (\ref{LT}), determining $T_e$ with good accuracy in quasi-1D samples, (\ref{Ly}).} {We have  confirmed the viability of the method performing measurements on graphene devices in the low-current regime, when $T_e\approx T$.
 For future applications, the method may be especially
useful for  graphene devices at higher
currents, where  overheating  is likely to arise from inefficient thermal
contact with the environment.}

\acknowledgements{ We acknowledge support from the EPSRC grant
EP/D031109, the Lancaster-EPSRC Portfolio Partnership
 EP/C511743, and ESF
 FoNE CRP ``SpiCo".}


\begin{references}

\bibitem{Wallace}
P.~R. Wallace, Phys. Rev. {\bf 71}, 622 (1947).

\bibitem{McClure}
J.~W. McClure, Phys. Rev. {\bf 104}, 666 (1956).

\bibitem{NovGeim}
A.~K. Geim and K.~S. Novoselov, Nat. Materials {\bf 6}, 183 (2007).

\bibitem{Geim:08}
S.~V. Morozov, K.~S. Novoselov, M.~I. Katsnelson, F.~Schedin, D.~C. Elias,
  J.~A. Jaszczak, and A.~K. Geim, Phys. Rev. Lett. {\bf 100}, 016602 (2008).

\bibitem{AAG}
I.~L. Aleiner, B.~L. Altshuler, and M.~E. Gershenson, Waves in Random Media
  {\bf 9}, 201 (1999); I.~L. Aleiner, Ya.~M. Blanter,  Phys. Rev.
 {\rm B} {\bf 65}, 115317 (2002).


\bibitem{Marcus:93}
C.~M. Marcus, R.~M. Westervelt, P.~F. Hopkins, and A.~C. Gossard, Phys. Rev.
 {\rm B} {\bf 48}, 2460 (1993).

\bibitem{McCann:06}
E.~McCann, K.~Kechedzhi, V.~I. Fal'ko, H.~Suzuura, T.~Ando, and B.~L.
  Altshuler, Phys. Rev. Lett. {\bf 97}, 146805 (2006);
K.~Kechedzhi, V.~I. Fal'ko, E.~McCann, and B.~L. Altshuler, Phys. Rev. Lett.
  {\bf 98}, 176806 (2007).

\bibitem{Sav:08}
F.~V. Tikhonenko, D.~W. Horsell, R.~V. Gorbachev, and A.~K. Savchenko, Phys.
  Rev. Lett. {\bf 100}, 056802 (2008).


\bibitem{AlKh:85}
B.~L. Altshuler and D.~E. Khmelnitskii,  JETP Lett.\  {\bf 42}, 359 (1985).

\bibitem{LSF}
P.~A. Lee, A.~D. Stone, and H.~Fukuyama, Phys. Rev. {\rm B} {\bf 35}, 1039
  (1987).

\bibitem{Kechedzhi:08}
K.~Kechedzhi, O.~Kashuba, and V.~I. Fal'ko, Phys. Rev. {\rm B} {\bf
77}, 193403 (2008).

\bibitem{Khar}
M.~Y. Kharitonov and K.~B. Efetov, Phys. Rev. {\rm B} {\bf 78},
033404 (2008).

\bibitem{Wurm} J.~Wurm, A.~Rycerz, I.~Adagideli, M.~Wimmer, K.~Richter,
H.U.~Baranger, arXiv:0808.1008.

\bibitem{Haug1}
T.~Schmidt, P.~K{\"o}nig, E.~McCann, V.~I. Fal'ko, and R.~J. Haug,
Phys. Rev.
  Lett. {\bf 86}, 276 (2001);
J.~K{\"o}nemann, P.~K{\"o}nig, T.~Schmidt, E.~McCann, V.~I. Fal'ko,
and R.~J. Haug,
  Phys. Rev. {\rm B} {\bf 64}, 155314 (2001).

\bibitem{anglbrckt}
In  measurements, $\langle \delta G
  \delta G'\rangle$
is the correlation function of the fluctuations of conductance
$\delta G(\varepsilon _{\text{F}})$ averaged over a wide range of
Fermi energies and magnetic fields. {For given $\varepsilon _\text{F} $, the averaging  over magnetic field is done by extracting from experimental data  the autocorrelator
$\langle \delta G
  \delta G'\rangle = \int_{B}^{B+B_0}\tfrac{\textrm{d}B'}{B_0}G(B',\varepsilon_{\text F})
G(B',\varepsilon_{\text F}+\Delta)$}.
In our theoretical analysis we use the standard substitution of such
an average  by the disorder ensemble averaging, based upon the
ergodicity hypothesis~\cite{AKL:86a}.


\bibitem{Tdistr}
In case of overheating
the temperature distribution in the sample may be inhomogeneous; then  Eq.~(\ref{dEc1})
determines the average  electron temperature in
the sample. 


\bibitem{diags}
Note in passing that only diagrams with two diffusion modes contribute in
the lowest order. The diagrams with three or four diffusion modes, taken into
account in \cite{LSF}, mutually cancel as they originate from the
disorder averaging of the $\mathcal{G}^R\mathcal{G}^R$ and
$\mathcal{G}^A \mathcal{G}^A$ contribution to the conductance  which
vanish in each given realization of disorder \cite{B+S} {($\mathcal{G}^{R/A}$ are the retarded/advanced Green's functions)}.


\bibitem{Geim0}
K.~S. Novoselov, A.~K. Geim, S.~V. Morozov, D.~Jiang, Y.~Zhang, S.~V. Dubonos,
  I.~V. Grigorieva, and A.~A. Firsov, Science {\bf 306}, 666 (2004).


\bibitem{efermi} {In graphene with the linear dispersion law $\varepsilon=vp$, this relation is $\varepsilon _\text{F} =2\hbar
v \sqrt{ \pi V_g C/e }$, where  $C$ is the
capacitance per unit area between the gate electrode and graphene.
For our geometry,  this relation reduces to
  $\varepsilon_\text{F} =30V_g^{1/2}$
 where $V_g$ is measured in Volts and $\varepsilon _\text{F} $ in meV.}

\bibitem{AKL:86a}
B.~L. Altshuler, V.~E. Kravtsov, and I.~V. Lerner,  JETP Lett.\  {\bf 43}, 441 (1986); O.~Tsyplyatyev, I.~L. Aleiner, V.~I.~Fal'ko, and I.~V. Lerner, Phys. Rev. {\rm
  B} {\bf 68}, 121301(R) (2003).

\bibitem{B+S} H.~U. Baranger and A.~D. Stone, Phys. Rev. {\rm B} {\bf 40}, 8169 (1989).

\end{references}

\end{document}